\documentclass[twocolumn,showpacs,floats,prl,aps]{revtex4}

\usepackage[dvips]{graphicx}
\usepackage{amsfonts}
\usepackage{amsmath}
\usepackage{amssymb}
\usepackage{exscale}

\begin{document}

\title{
	   Two-Fermi-surface superconducting state and a nodal $d$-wave gap 
	   in the electron-doped Sm$_{1.85}$Ce$_{0.15}$CuO$_{4-\delta}$ cuprate superconductor.
	   }
\author{A.~F.~Santander-Syro,$^{1,2}$
		M.~Ikeda,$^{3}$
		T.~Yoshida,$^{3}$ 
		A.~Fujimori,$^{3}$
		K.~Ishizaka,$^{4}$
		M.~Okawa,$^{4}$
		S.~Shin,$^{4}$
		R.~L.~Greene,$^{5}$ and
		N.~Bontemps$^{2}$. 
		}
\address{$^1$CSNSM, Universit\'e Paris-Sud and CNRS/IN2P3, B\^atiments 104 et 108, 91405 Orsay cedex, France.}
\address{$^2$Laboratoire Physique et Etude des Mat\'eriaux, UPR-5 CNRS, ESPCI, 10 rue Vauquelin, 75231 Paris cedex 5, France.}
\address{$^3$Department of Physics, University of Tokyo, Hongo, Bunkyo-ku, Tokyo 113-0033, Japan.}
\address{$^4$Institute for Solid State Physics, University of Tokyo, Kashiwa, Chiba 277-8581, Japan.}
\address{$^5$Center for Nanophysics and Advanced Materials, Department of Physics, 
		University of Maryland, College Park, MD 20742.}
\date{\today}

\begin{abstract}
We report on laser-excited angle-resolved photoemission spectroscopy (ARPES) in the electron-doped cuprate 
Sm$_{1.85}$Ce$_{0.15}$CuO$_{4-\delta}$. 
The data show the existence of a nodal hole-pocket Fermi-surface
both in the normal and superconducting states.
We prove that its origin is long-range antiferromagnetism by an analysis of the coherence factors
in the main and folded bands.
This coexistence of long-range antiferromagnetism and superconductivity implies that electron-doped cuprates 
are two-Fermi-surface superconductors.
The measured superconducting gap in the nodal hole-pocket is compatible with a $d$-wave symmetry.
\end{abstract}

\pacs{74.25.Jb, 74.72.Ek}

\maketitle

In cuprates, the ground state at half-filling is an antiferromagnetic (AFM) insulator,
and superconductivity appears upon doping~\cite{Tokura-AFM-SC-Competition-eDoped}.
In the electron-doped cuprates, the superconducting (SC) and N\'eel temperatures, as well as the actual carrier density,
depend on both the Ce ($x$) and O-vacancy ($\delta$) concentrations~\cite{Kang-Annealing-eDoped}. 
In these materials, transport studies~\cite{Greene-QCP-PCCO, Greene-MagnetoTransport-PCCO}
suggested a coexistence of antiferromagnetism and superconductivity
up to dopings $x \approx 0.16$, slightly above the optimal doping ($x = 0.15$)
On the other hand, inelastic neutron scattering experiments claimed that long-range antiferromagnetism 
and superconductivity are adjacent, rather than coexisting, phases~\cite{Motoyama-NeutronsNCCO}. 
These apparently contradictory results might be related to different annealing conditions, as stated above, 
or sample inhomogeneity. 

Previous ARPES experiments in reduced SC samples 
reported band folding~\cite{Ikeda-LnCeCuO-PRB, Matsui-BandFolding-UndNCCO, Park-UndSCCO},
interpreted as either due to an anisotropic AFM gap~\cite{Matsui-BandFolding-UndNCCO}
or to short-range-order antiferromagnetism~\cite{Park-UndSCCO}.
A Fermi-surface electron-pocket around $(0, \pi/a)$ ($a$ is the in-plane lattice parameter)
was observed in underdoped or optimally-doped samples~\cite{Ikeda-LnCeCuO-PRB, Park-UndSCCO}.
Other ARPES~\cite{Matsui-SCgap} and Raman~\cite{Blumberg-Raman-SCgap} experiments in optimally
doped samples claimed a continuous `non-monotonic $d$-wave' SC gap,
with the gap increasing from zero at $(\pi/a, \pi/a)$ to a maximum near the `hot-spots', 
where the Fermi surface crosses the AFM zone-boundary, and stalling at a nearly constant amplitude 
in the region between the hot spots and $(0, \pi/a)$.
Recently, Shubnikov-de Haas (SdH) oscillations in SC samples of Nd$_{2-x}$Ce$_{x}$CuO$_{4-\delta}$ 
showed the existence of small Fermi-surface pockets of area $\sim 1.1$\% of the Brillouin zone
in optimally- and slightly over-doped samples, interpreted as the nodal hole pockets created by
long-range AFM ordering~\cite{Helm-SdH-NCCO}. 
However, the crucial issues of proving directly the coexistence 
of long-range antiferromagnetism and superconductivity,
the existence of a nodal hole-pocket Fermi-surface, 
and how the SC gap opens over the AFM-folded band structure, remain open.

%
Laser-excited ARPES is essential to address these questions. 
First, electron-doped cuprates do not present a unique terminal surface 
upon cleaving, because of their crystal structure. This calls for a bulk sensitive technique. 
Second, their nodal Fermi wave-vector is very close 
to the AFM zone boundary, so that any folded bands are very
close in momentum space --besides having weak spectral weight and large linewidths~\cite{Armitage-SelfEnNCCO}.
This needs high momentum resolution and low background from inelastically-scattered electrons.
Third, the expected SC gap is of the order of a few meV,
requiring high energy resolution.
Laser-ARPES gathers all these advantages, due to the low photoelectron kinetic energy and the high
intrinsic resolution of the laser light~\cite{Kiss-BulkSensitive-SrVO3, Koralek-LaserARPES-Bi2212}.

In this Letter we show, using laser-ARPES, 
that long-range antiferromagnetism and superconductivity coexist 
in the electron-doped cuprate Sm$_{1.85}$Ce$_{0.15}$CuO$_{4-\delta}$ (SCCO).
The data reveal, for the first time, a hole-pocket Fermi-surface around $(\pi/2a, \pi/2a)$.
The spectral weights in the `main' and `folded' bands of this hole-pocket
coincide with the coherence factors of a system with homogeneous long-range antiferromagnetism.
The SC gap observed in the hole pocket is $d$-wave like.
Our data imply that electron-doped cuprates are two-Fermi-surface superconductors.

High-quality single crystals of SCCO were grown by a flux method and
then annealed under low-oxygen pressure to render them SC~\cite{Greene-XtalGrowth} 
with a $T_c = 12.5$~K determined by SQUID magnetometry.
Wavelength dispersive X-ray analysis yielded $x=0.15 \pm 0.01$.
The high-resolution laser-ARPES experiments were performed with a Gammadata R4000 analyzer and an ultraviolet laser
at $h\nu = 6.994$~eV~\cite{Kiss-LaserARPES}. 
The energy and momentum resolutions were respectively 2~meV and $3\times 10^{-3}$~\AA$^{-1}$.
The pressure of the chamber was $\lesssim 5\times10^{-11}$~Torr throughout all the measurements.
The Laue-oriented crystals were cleaved {\it in situ} at $5.7$~K, just before the measurements.
The doping was checked by a Fermi-surface fit to the experimental
Fermi momenta in the nodal arc using a single-band tight-binding model~\cite{Optics-PCCO-Millis-Zimmers},
giving $n_e=0.14 \pm 0.01$ per formula unit. 
 
We recall that ARPES measures the occupied part of the energy ($\omega$) and momentum ($\mathbf{k}$) dependent
single-particle spectral function $A(\mathbf{k},\omega)$.
We note $\epsilon_{\mathbf{k}}$ the independent-particle dispersion, 
$\mathbf{Q} = (\pi/a, \pi/a)$ the folding vector, and $\Delta$ the AFM gap at $\mathbf{k}=\mathbf{Q}$.  
Then, the conduction ($E_{\mathbf{k}}^{c}$, $+$) and valence ($E_{\mathbf{k}}^{v}$, $-$) bands 
in the presence of long-range AFM order
are~\cite{Bansil-TwoBand-NonMonoDwave, Belen-Leni-Uk2Vk2}:
\begin{equation}
	E_{\mathbf{k}}^{c,v} = 
	\frac{1}{2} \left[ \epsilon_{\mathbf{k}}+\epsilon_{\mathbf{k+Q}} 
	\pm E_{\mathbf{k}}^{0} \right],
\label{EqBandFolding}
\end{equation}
with
\begin{equation}
	E_{\mathbf{k}}^{0} = \sqrt{\left( \epsilon_{\mathbf{k}}-\epsilon_{\mathbf{k+Q}} \right)^{2} 
	                    +4\Delta^{2}}.
\label{EqFoldingGap}	                    
\end{equation}
The spectral function is~\cite{Bansil-TwoBand-NonMonoDwave, Belen-Leni-Uk2Vk2}:
%
\begin{equation}
	A(\mathbf{k},\omega) = \frac{1}{\pi} 
	\left[ 
	\frac{\Gamma U_{\mathbf{k}}^{2}}{\left( \omega - E_{\mathbf{k}}^{c} \right)^{2} + \Gamma^{2}}
	+ \frac{\Gamma V_{\mathbf{k}}^{2}}{\left( \omega - E_{\mathbf{k}}^{v} \right)^{2} + \Gamma^{2}}
	\right],
\label{EqSpectralFn}
\end{equation}
where $\Gamma$ is the electron scattering rate. 
The coherence factors, which quantify the removal probability of an electron from the `main' and `folded' bands, 
are~\cite{Bansil-TwoBand-NonMonoDwave, Belen-Leni-Uk2Vk2}:
\begin{equation} 
	U_{\mathbf{k}}^{2} = \frac{1}{2} \left[ 1 + \frac{\epsilon_{\mathbf{k}}-\epsilon_{\mathbf{k+Q}}} 
							 {E_{\mathbf{k}}^{0}} \right] \text{, } 
	V_{\mathbf{k}}^{2} = \frac{1}{2} \left[ 1 - \frac{\epsilon_{\mathbf{k}}-\epsilon_{\mathbf{k+Q}}} 
							 {E_{\mathbf{k}}^{0}} \right]. 
\label{EqUkVk}
\end{equation}

Experimentally, we only have access to features below $E_F$, namely $E_{\mathbf{k}}^{v}$ and $V_{\mathbf{k}}^{2}$.
However, for $k \parallel k_{x}$ at $k_{y}=\pi/2a$, it can be shown that
$V^{2}(\pi/2a + k, \pi/2a) = U^{2}(\pi/2a - k, \pi/2a)$.
In this case, the spectral weights of the main and folded bands {\it below} $E_F$
correspond to $V_{\mathbf{k}}^{2}$ and $U_{\mathbf{k}}^{2}$, respectively, verifying the sum rule
$U_{\mathbf{k}}^{2} + V_{\mathbf{k}}^{2} = 1$. 

\begin{figure}[tbhp]
  \begin{center}
     \includegraphics[width=8cm]{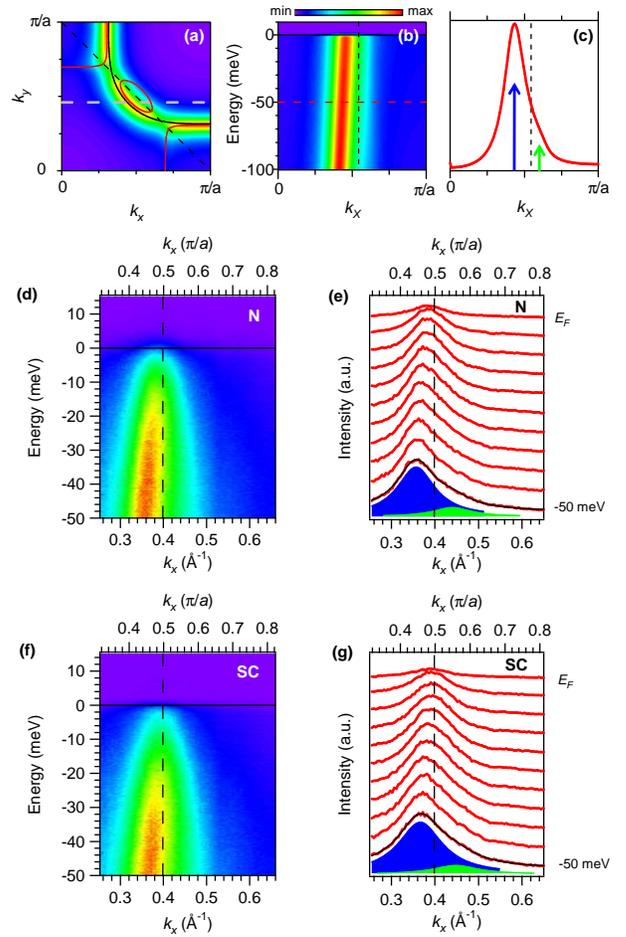}
  \end{center}
  \caption{\label{Fig1}
		    (a-c) Simulation of an optimally-electron-doped cuprate~\cite{Optics-PCCO-Millis-Zimmers},  
		    with an AFM gap of 150~meV, and a scattering rate of 200~meV.  
		    Panel (a) shows the Fermi surface without and with folding 
		   	(black and red lines, respectively), 
		   	and spectral function at $E_{F}$ (color plot). 
		   	The dashed gray line represents a cut parallel to $k_x$ passing through the node,
		   	as in our experiments.
		   	(b) Energy-momentum map along the dashed gray line in (a). 
		   	(c) MDC along the red line in (b) at $E=-50$~meV. 
		   	The main and folded bands are shown with the blue and green arrows, respectively.
		   	(d-g) Experimental energy-momentum maps and momentum distribution curves 
		    in the normal state (N) at $20$~K [(d) and (e)] and in the
		    SC state at $5.7$~K [(f) and (g)].  
		    Panels (e) and (g) display a fit to the MDC at $E=-50$~meV
		    using two Lorentzians of the same width for the main (blue) and folded (green) bands.
		    The resulting fit function is the black
		    curve on top of the data. The dashed blue and green lines are guides to the eye,
		    to show the dispersion of the main and folded bands.
		    In all figures, the AFM zone boundary is shown by a black dashed line.
		   }
\end{figure}
Figures.~\ref{Fig1}(a-c) are a guide to our data analysis. They present a simulation of AFM folding in the 
spectral function of an electron-doped cuprate
using values for the AFM gap and scattering rate close to the ones 
inferred from our data (see later). 
Due to coherence factors and the scattering introduced in the simulation, 
the folded band is seen neither in the Fermi surface shown in figure~\ref{Fig1}(a)
nor in the energy-momentum map along the node shown in figure~\ref{Fig1}(b). However, its presence yields a shoulder 
(or an overall assymetric line-shape) in the momentum distribution curves (MDCs), as in figure~\ref{Fig1}(c), 
and observed in the data discussed below.

Figures~\ref{Fig1}(d-g) show measurements for cuts along $k_{x}$ passing through the node.
Figs.~\ref{Fig1}(d, e) present the measured energy-momentum intensity maps and the corresponding MDCs,
in the normal state at $20$~K [to be compared to figures~\ref{Fig1}(b)-(c)].
Figs.~\ref{Fig1}(f, g) display the equivalent data in the SC state at $5.7$~K. 
Note that, as in the simulations, the MDCs in both the normal and SC states
present an asymmetric shape, with a distinct peak at wavevectors smaller than 
the AFM zone boundary ($k_{AF}$, black dashed line) 
and a shoulder at $k_{x}>k_{AF}$. 
The peak and shoulder disperse towards each other as they approach $E_F$. 
We found that the MDCs can be very well fitted by two Lorentzians symmetrically distributed around $k_{AF}$, 
of equal width (full width at half maximum $\sim 0.1$~\AA$^{-1}$) but different amplitudes, 
as shown by the fits to the MDCs at $E=-50$~meV in Figs.\ref{Fig1}(e, g)
(we checked that relaxing all these conditions does not change the results that follow).
This is fully compatible with the simulations shown in Fig.~\ref{Fig1}(c). 
Therefore, we ascribe the peak and
shoulder features to the main and folded bands at the AFM wave vector, respectively. 
To prove that such an assignment is correct, 
we determined the coherence factors for the main and folded bands in the normal state 
by using the two independent methods described below.
%

\begin{figure}[t]
  \begin{center}
     \includegraphics[width=8cm]{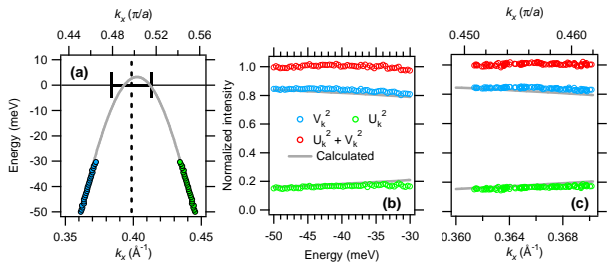}
  \end{center}
  \caption{\label{Fig3}
		   (a) Dispersion of the main and folded bands 
		   (blue and green filled circles, respectively) 
		   and fit with a tight binding model with an AFM
		   gap of 130~meV (gray curve). 
		   The vertical dashed line shows the position of the AFM zone boundary
		   deduced from the experimental geometry 
		   with its associated error due to a possible in-plane sample misalignment smaller than $3^{\circ}$.  
		   (b, c) Spectral weight of the main and folded bands, $V_{k}^{2}$ and $U_{k}^{2}$
		   (blue and green open circles, respectively), total spectral weight (red open circles),
		   and coherence factors calculated from the fit in panel (a) (gray curves), 
		   plotted as a function of energy in (b), and as a function of momentum in (c).
		   In all panels, the error bars are of the symbols' size.
		   }
\end{figure}
For the first method, we performed the above two-Lorentzian fits for the MDCs in the range 
$-50$~meV~$\leq E\lesssim -30$~meV, because for energies closer to $E_F$ 
the fit breaks down.  
The widths of the Lorentzians were found to be constant within experimental uncertainties in this energy range.
Thus, the amplitude parameter of each Lorentzian represents
$V_{\mathbf{k}}^{2}$ and $U_{\mathbf{k}}^{2}$ for the main and folded bands, respectively.
The dispersions for the main and folded bands, resulting from the centers of the Lorentzians,
are shown by the blue and green filled circles in Fig.~\ref{Fig3}(a). 
The corresponding spectral weights are represented as a function of energy
by the blue and green open circles in Fig.~\ref{Fig3}(b), and as a function of momentum in Fig.~\ref{Fig3}(c).
The total spectral weight $U_{\mathbf{k}}^{2} + V_{\mathbf{k}}^{2}$,
shown by the red open circles in Figs.~\ref{Fig3}(b)-(c),
is found to be constant within error bars. This indicates that the sum rule is obeyed. 
Thus, the average of the total spectral weight was taken   
as an absolute normalization factor for the amplitude of each band
to define the vertical scales in Figs.~\ref{Fig3}(b)-(c).
Note that the experimental band velocity is about 20~meV$/$0.01~\AA$^{-1} = 2$~eV~\AA,
which together with the linewidth of $\sim 0.1$~\AA$^{-1}$ from Fig.~\ref{Fig1}(e) gives a 
scattering rate of approximately 200~meV, as in the simulations presented in Fig.~\ref{Fig1}.

For the second method, we calculated $V_{\mathbf{k}}^{2}$ and $U_{\mathbf{k}}^{2}$ from
Eqns.~\ref{EqBandFolding}, \ref{EqFoldingGap}, and~\ref{EqUkVk}.
These are completely determined by a model for the unfolded band, 
the folding vector and the AFM gap, {\it independently} of the experimental coherence factors.
We fitted the experimental main and folded bands in Fig.~\ref{Fig3}(a)
using a tight-binding band for $\epsilon_{\mathbf{k}}$ 
(we checked that other model bands, like a linearly-dispersing band, give consistent results)
folded by a vector $\mathbf{Q} = (\pi/a, \pi/a)$
with an AFM gap $\Delta = 130$~meV --compatible with other estimates 
for samples of similar doping~\cite{Park-UndSCCO,Optics-PCCO-SDWgap-Zimmers,Zimmers-STM-SCCO}.
The resulting calculated dispersion and coherence factors are represented by the grey lines in Figs.~\ref{Fig3}(a)-(c).
There is an excellent agreement between the two sets of coherence factors determined from the two independent procedures. 
This is unambiguous evidence for long-range AFM band-folding 
and the presence of a small nodal hole-pocket Fermi-surface as shown in Fig.~\ref{Fig3}(a).
In fact, note that a hypothetical superposition of AFM non-SC regions and paramagnetic SC regions
(which would be the case for, {\it e.g.}, a non-homogeneous sample or for short-range AFM order), or equivalently, 
a superposition of single-sheet and two-sheet Fermi surfaces cannot possibly explain our observations: 
the summed spectral weights of the superposed bands would obliterate any correlation 
to the expected coherence factors~\cite{Park-UndSCCO}. 
Additionally, scanning tunneling experiments on the same batch of samples 
found evidence for a local coexistence of antiferromagnetism and superconductivity 
in all the explored regions of the sample~\cite{Zimmers-STM-SCCO}. 
We checked (not shown) that similar analysis can be performed for band-folding observed  
along other cuts off the node, reinforcing the robustness of our conclusions. 

We can now estimate a lower bound for the AFM correlation length ($\xi_{AFM}$)
from the observed AFM folding:   
it was shown that dispersive folded bands exist  
when $\xi_{AFM} \gg \hbar v_{F}/(\pi k_{B}T)$, where $v_F$ is the Fermi velocity and the right-hand-side
is the electron thermal wavelength~\cite{Lee-Rice-Anderson-FluctuationsPierls, Vilk-FoldedBands-AFMCorrLength}.
From the measured $\hbar v_{F} \approx 2$~eV~\AA, we deduce that at 5~K, in the SC state,
$\xi_{AFM} \gg 1480$~\AA, or equivalently $\xi_{AFM} \gg 400 a$,
in agreement with estimates from recent SdH oscillations~\cite{Helm-SdH-NCCO}.
Furthermore, according to neutron scattering data in other electron-doped cuprates, 
the onset of long-range AFM order is reached 
when $\xi_{AFM}/a \sim 200-400$~\cite{Motoyama-NeutronsNCCO, Greven-SpinCorrelations-AFM-NonSC-NCCO}. 
Hence, we conclude that long-range antiferromagnetism and superconductivity coexist in the samples studied here.

\begin{figure}[t]
  \begin{center}
     \includegraphics[width=8cm]{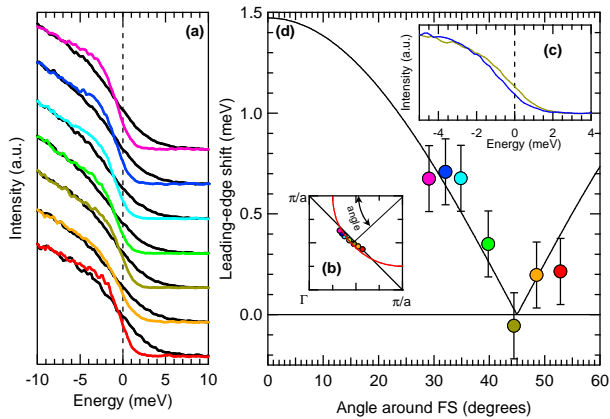}
  \end{center}
  \caption{\label{Fig4}
		   (a) EDCs integrated over $\pm 0.05$~\AA$^{-1}$
		   around $k_F$ in the normal state at $20$~K (black lines) and in the SC
		   state at $5.7$~K (color lines). 
		   The different colors represent 
		   different loci around the nodal arc region of the Fermi surface --see (b), 
		   where the angle around the Fermi surface (FS) is also defined.
		   (c) Enlarged view of the two EDCs in the SC state at the node and 
		   at $\sim 12^{\circ}$ away. 
		   (d) Leading-edge shift (SC with respect to normal) 
		   of the EDCs at $k_F$ shown in panel (a), as a function of the angle.
		   The black line is a fit to the data
		   using a monotonic $d$-wave gap of amplitude $\sim 1.5$~meV.
		   }
\end{figure}
We now study the SC gap in the hole pocket around the diagonal direction.
Figure~\ref{Fig4}(a) presents the energy distribution curves (EDCs) at the Fermi wave vector $k_F$
in the normal state at $20$~K (black lines) and the SC state at $5.7$~K (color lines)
at several loci in the Fermi-surface hole-pocket specified in Fig.~\ref{Fig4}(b).
Fig.~\ref{Fig4}(c) shows the two EDCs at 5.7~K along the node and $\sim 12^{\circ}$ away from the node. 
In the SC state, the leading edge of each EDC shifts to larger binding energies as one moves away from the node. 
In order to estimate the momentum-dependence of the SC gap ($\Delta_{SC}$),
we fitted the EDCs in the normal and SC states in figure~\ref{Fig4}(a) by resolution-broadened 
Fermi-Dirac functions with the Fermi energy (defining the leading-edge energy) 
being determined by the fit.
The resulting leading-edge shift ($\Delta_{LE}$) as a function of angle around the Fermi-surface 
is shown in Fig.~\ref{Fig4}(d). 
It is compatible with a $d$-wave gap 
of amplitude $\sim 1.5$~meV, as shown by the fit (black line) in Fig.~\ref{Fig4}(d).
Using $T_c = 12.5$~K, and the empirical relation $2\times \Delta_{LE} \approx \Delta_{SC}$~\cite{Matsui-SCgap}, 
the maximum SC gap that the system would have 
in absence of band folding is $\Delta_{SC}^{max}/k_B T_c \approx 2.8$.
This is comparable to the weak-coupling $d$-wave BCS value of $2.14$
and to the ratio of $2.2$ from measurements of the SC gap near the hot spots 
in other electron-doped cuprates~\cite{Matsui-SCgap, Blumberg-Raman-SCgap}.

In summary, our data give new crucial inputs to the understanding of 
electron-doped cuprates.
We directly observed the small Fermi-surface nodal hole pockets, where a $d$-wave-like SC gap opens,
and proved that long-range antiferromagnetism and superconductivity coexist in optimally-doped SCCO.
Together with the previously reported electron pockets at $(0, \pi/a)$~\cite{Ikeda-LnCeCuO-PRB, Park-UndSCCO}
and the non-monotonic SC gap over the whole Fermi surface~\cite{Matsui-SCgap, Blumberg-Raman-SCgap},
our observations imply that electron-doped cuprates are two-Fermi-surface superconductors,
and suggest that, in contrast to the nodal hole pocket, the electron pocket around $(0, \pi/a)$ 
becomes fully gapped in the SC state, the gap being nearly isotropic.
As proposed by several theories 
of electron-doped cuprates~\cite{Yuan-Ting-TwoBand-NonMonoDwave, Bansil-TwoBand-NonMonoDwave, Luo-Xiang-SFD},
a two-SC-gap picture would explain the apparent non-monotonic $d$-wave gap
and the anomalous temperature dependence of the superfluid density~\cite{Kim-Anomalous-SFD},
which presents a positive curvature close to $T_c$ --a property of two-band superconductors~\cite{Xiang-Wheatley-TwoBand-SCs}.

We thank B. Liang and P. Li for help in the sample growth and characterization, 
and B.~Valenzuela, E.~Bascones, A.~Millis and A.-M.~Tremblay for discussions.
Work at University of Tokyo was supported by a Grant-in-Aid for Priority Area ``Invention 
of Anomalous Quantum Materials" from MEXT, Japan.
Work at the University of Maryland was supported by NSF DMR-0653535.

\end{document}